\documentclass[preprint,showpacs,preprintnumbers,amsmath,amssymb]{revtex4-1}
\usepackage{graphicx}
\usepackage{subfigure}
\usepackage{dcolumn}
\usepackage{bm}
\usepackage{amssymb}

\def \un{{\bf{\hat n}}}
\def \uk{{\bf{\hat k}}}
\def \prd{{\it Phys. Rev. D, }}

\def \apj{{\it Ap. J., }}
\def \GRG{{\it Gen. Relativ. Gravit., }}

\begin{document}
\title{Nano-Hertz Gravitational Waves Searches with \\
  Interferometric Pulsar Timing Experiments}

\author{Massimo Tinto}
\affiliation{Jet Propulsion Laboratory,\\
  California Institute of Technology, \\
Pasadena, CA 91109}
\email{Massimo.Tinto@jpl.nasa.gov}

\date{\today}

\begin{abstract}
  We estimate the sensitivity to nano-Hertz gravitational waves of
  pulsar timing experiments in which two highly-stable millisecond
  pulsars are tracked simultaneously with two neighboring radio
  telescopes that are referenced to the same time-keeping subsystem
  (i.e. ``the clock'').  By taking the difference of the two
  time-of-arrival residual data streams we can exactly cancel the
  clock noise in the combined data set, thereby enhancing the
  sensitivity to gravitational waves. We estimate that, in the band
  ($10^{-9} - 10^{-8}$) Hz, this ``interferometric'' pulsar timing
  technique can potentially improve the sensitivity to gravitational
  radiation by almost two orders of magnitude over that of
  single-telescopes.  Interferometric pulsar timing experiments could
  be performed with neighboring pairs of antennas of the forthcoming
  large arraying projects.
\end{abstract}

\maketitle

The direct detection of gravitational waves is one of the most
challenging experimental efforts in physics today. A successful
observation will not only represent a great triumph in experimental
physics, but will also provide a new observational tool for obtaining
a better and deeper understanding about their sources, as well as a
unique test of the various proposed relativistic metric theories of
gravity \cite{Thorne1987}.

Pulsar timing experiments aimed at the detection of gravitational
radiation have been performed for decades now.  The basic principle
underlining the pulsar timing technique is the same as that of other
gravitational wave detector designs
\cite{LIGO,VIRGO,GEO,TAMA,DOPPLER,PPA98}: to monitor the frequency
variations of a coherent electromagnetic signal exchanged by two or
more ``point particles'' separated in space. As a pulsar continuously
emits a series of radio pulses that are received at Earth, a
gravitational wave passing across the pulsar-Earth link introduces
fluctuations in the time-of-arrival (TOA) of the received
electromagnetic pulses. By comparing the pulses TOAs against those
predicted by a model, it is in principle possible to detect the
effects induced by any time-variable gravitational fields present,
such as the transverse-traceless metric curvature of a passing plane
gravitational wave train \cite{Pulsar}. The frequency band in which
the pulsar timing technique is most sensitive to ranges from about
$10^{-9}$ to $10^{-6}$ Hz, with the lower-limit essentially determined
by the overall duration of the experiment ($10^{-9} \ {\rm Hz} \simeq
1/{\rm 30 \ years}$), and the upper limit identified by the
signal-to-noise ratio of the received radio pulses. To attempt
observations of gravitational waves in this way, it is thus necessary
to control, monitor and minimize the effects of other sources of
timing fluctuations, and, in the data analysis, to use optimal
algorithms based on the different characteristics of the pulsar timing
response to gravitational waves (the signal) and to the other sources
of timing fluctuations (the noise).

A quantitative analysis of the noise sources affecting millisecond
pulsar timing searches for gravitational radiation has been
highlighted in a recent paper by Jenet {\it et al.}  \cite{JAT2011}.
There it has been shown that, in the region ($10^{-9} - 10^{-8}$) Hz of
the accessible frequency band, the main noises are due to:
\begin{enumerate}
\item finiteness of the signal-to-noise ratio in the raw
  observations;
\item uncertainties in solar system ephemerides, which are used to
  correct arrival times at the Earth to the barycenter of the solar
  system;
\item variation of the index of refraction in the interstellar and
  interplanetary plasma; 
\item intrinsic rotational stability of the pulsar, and
\item instability of the local clock against which pulsars are timed,
  and noise in time transfer if the clock is not located at the
  observatory site.
\end{enumerate}
Although the timing fluctuations induced by most of these noises can
be in principle either reduced or calibrated out, the fundamental
noise-limiting sensitivity of pulsar timing experiments is imposed by
the timing-fluctuations inherent to the reference clocks that control
the TOA measurements.  The noise analysis in \cite{JAT2011} indicated
that clocks such as the Linear-Ion-Trapped-Standard (LITS) (presently
in-the-field state-of-the art atomic clocks with long-time-scale
timing stability) would result in a sinusoidal strain sensitivity of
pulsar timing searches for gravitational waves to a level of about
$10^{-15}$ after coherently integrating the data for a period of $10$
years. Since the characteristic wave amplitude associated with a
super-massive black-hole-binaries background is predicted to be of
comparable magnitude at the frequency $3 \times 10^{-9}$ Hz
\cite{Black1,Black2}, it is clear that a single telescope will not be
able to unambiguously detect such a gravitational wave signal.

A method for statistically enhancing the signal-to-noise ratio of
pulsar timing experiments to backgrounds of gravitational waves was
first proposed by Hellings and Downs \cite{HD1983} and further
analyzed and improved by Jenet {\it et al.}  \cite{JHLM2005}. This
technique relies on cross-correlating pairs of TOA residuals data
taken by an array of radio telescopes observing highly-stable
millisecond pulsars.  Since the gravitational wave background is
common to all the timing residuals while many noises affecting them are
in principle uncorrelated, the cross-correlation technique enhances
the strength of the gravitational wave signal over that of the noise.

The experimental configuration we propose in this letter is
complementary to the cross-correlation method, in that it provides a
way for exactly canceling the clock noise from pairs of TOAs
residuals.  This is possible by relying on TOAs residuals measured
simultaneously by two neighboring radio telescopes tracking, with the
same clock, two highly-stable millisecond pulsars located in two
different parts of the sky. Since the two TOAs responses to the
gravitational wave signal will be different while the clock timing
noise transfer function in the two TOA residuals will be the same, by
taking the difference of the two time series the clock noise is
exactly canceled from the resulting synthesized data while
sensitivity to a gravitational wave signal is preserved.

\begin{figure}
\centering
\includegraphics[width=6.5in]{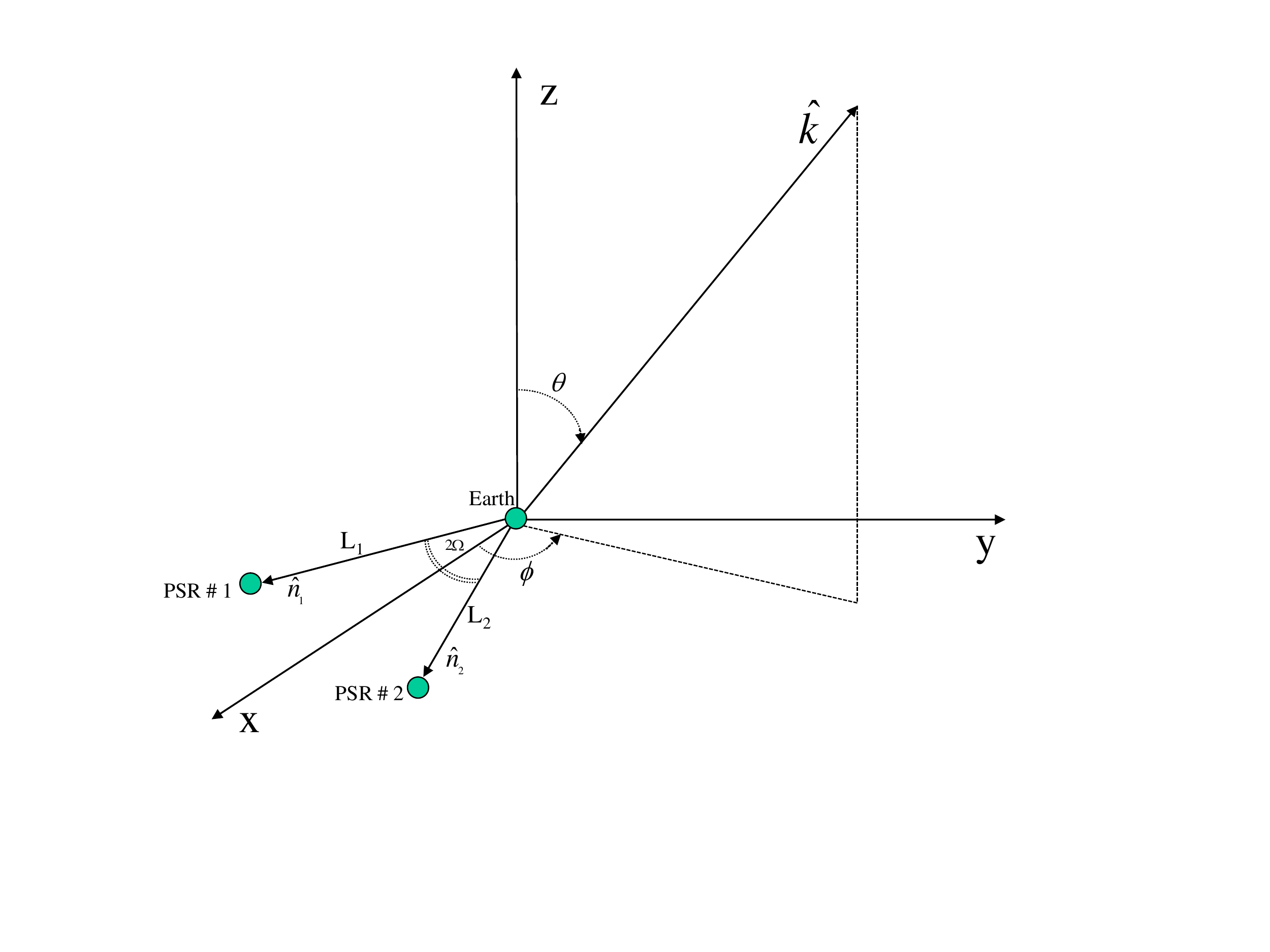}
\caption{Two millisecond pulsars are tracked simultaneously by two radio telescopes
  sited near each other. The unit vector $\uk$ defines the direction
  of propagation of the gravitational wave signal and ($\theta, \phi$)
  are the usual polar angles associated with it. The two unit vectors,
  ($\un_1, \un_2$), define the directions in the sky to the two
  pulsars, and the angle between them has been denoted with $2\Omega$.
  The bisector of this angle coincides with the $x$-axis of an
  orthogonal coordinate system in which the $y$-axis lies in the plane
  defined by the two unit-vectors ($\un_1, \un_2$), and the $z$-axis
  is chosen to form a right-handed orthogonal triplet ($x, y, z$).}
\label{fig1}
\end{figure}

Let us denote with $R_1 (t)$, $R_2(t)$ the two time-of-arrival
residuals measured at time $t$ at the two telescopes. Their time
derivatives can be written in the following way
\cite{EW1975,Burke1975,Wahlquist1987}
\begin{eqnarray}
\frac{d R_i (t)}{d t} & = & (1 - \uk \cdot \un_i) \ [\Psi_i(t - L_i(1 + \uk \cdot \un_i)) - \Psi_i (t)] 
\nonumber
\\
& & + C (t) + N_i (t) \ \ \ \ i = 1, 2 \ ,
\label{Residuals}
\end{eqnarray}
where we have denoted with $C (t)$ the random process associated with
frequency fluctuations due to the clock (common to the two timing
residual measurements), with $ N_i (t) \ , \ i = 1, 2$ the two random
processes associated with frequency fluctuations due to all other
noise sources affecting the timing residuals, and the function $\Psi_i
(t)$ contains the contribution of a spin-2 gravitational wave
transverse-traceless strain tensor ${\bf {h}}$ in the following way
\begin{equation}
\Psi_i (t) = \frac{(\un_i \cdot {\bf {h}} \cdot \un_i)}{2[1 - (\uk
\cdot \un_i)^2]} \ .
\label{psi}
\end{equation}
In Eqs.(\ref{Residuals}, \ref{psi}) $\ \ \un_i L_i \ , \ i = 1, 2$ is
the vector oriented from telescope $i$ to pulsar $i$, $\uk$ is the
unit vector associated with the direction of propagation of the wave,
and the symbol $\cdot$ is the operation of scalar product (see Fig.
\ref{fig1}). If we denote with $\Delta {\dot R}$ the difference of the
time-derivative of the two timing residuals given in Eq.
(\ref{Residuals}), we find the resulting interferometric pulsar timing
response
\begin{eqnarray}
\Delta {\dot R} (t) & \equiv & {\dot R}_1 (t) - {\dot R}_2 (t) = N_1 (t) - N_2 (t) 
\nonumber
\\
& + & [(1 - \uk \cdot \un_2) \ \Psi_2(t) - (1 - \uk \cdot \un_1) \
\Psi_1(t)] 
\nonumber
\\
& + & [(1 - \uk \cdot \un_1) \Psi_1(t - L_1(1 + \uk \cdot \un_1)) 
\nonumber
\\
& - &
(1 - \uk \cdot \un_2) \Psi_2(t - L_2(1 + \uk \cdot \un_2))] \ .
\label{Interfero}
\end{eqnarray}

Note how the {\it two-pulse} time-structure of the time-derivative of
the two TOA residuals given in Eq. (\ref{Residuals}) becomes a {\it
  three-pulse} response in the interferometric pulsar timing
expression given by Eq. (\ref{Interfero}); the three pulses
correspond to the times a gravitational wave interacts with the Earth
(which are the same in the two TOAs residuals) and the two millisecond
pulsars.

To assess the effectiveness of the interferometric pulsar-timing
technique we have estimated its sensitivity by relying on the
noise-model discussed in \cite{JAT2011}. In particular, we have
assumed that: (i) multiple-frequency measurements can be implemented
in order to adequately calibrate timing fluctuations due to the
intergalactic and interplanetary plasma; (ii) more accurate solar
system ephemerides will be available by the time these experiments
will be performed; and (iii) the tracked millisecond pulsars have
frequency stabilities better than those of the operational ground
clocks.  A recent stability analysis of presently known millisecond
pulsars \cite{Verbiest2009} has shown that there might exist some with
frequency stabilities superior to those displayed by the most stable
operational clocks in the ($10^{-9} - 10^{-8}$) Hz frequency band.  It
should be emphasized, however, that the interferometric pulsar timing
experiments discussed in this letter can also be used to isolate and
study the intrinsic pulsars spin-noises with a much higher precision
than what currently achievable by single-antenna experiments.

The sensitivity of a detector of gravitational waves has been
traditionally taken to be equal to (on average over the sky and
polarization states) the strength of a sinusoidal gravitational wave
required to achieve a given signal-to-noise ratio (SNR) over a
specified integration time, as a function of Fourier frequency. Here
we will assume a ${\rm SNR} = 1$ over an integration time of $10$
years. In order to compute the interferometric pulsar timing
sensitivity we will consider two millisecond pulsars of comparable
intrinsic stability and at a distance of $1$ kpc from the Earth. We
will also assume the following expression for the power spectrum of
the noises $N_i \ , \ i = 1, 2$ introduced in Eq.  (\ref{Residuals})
\begin{equation}
S_{N} (f) = 3.4 \times 10^{-8} \ f^2 \ {\rm Hz}^{-1} \ ,
\label{Sy}
\end{equation}
which corresponds to a white-timing noise of $100$ nsec.
\cite{JAT2011} in a Fourier band +/- 0.5 cycles/day (i.e. one sample
per day). The $100$ nsec. level is the current timing goal of leading
timing array experiments as three pulsars are being timed to this
level \cite{Verbiest2009}.

The gravitational wave sensitivity of our proposed interferometric
pulsar timing technique is then defined as the wave amplitude required
to achieve a signal-to-noise ratio of $1$ in a ten years integration
time: $\sqrt{2 S_N (f) \ B}$/(root-mean-squared gravitational wave
response). Note the factor $2$ multiplying the spectrum $S_N$ follows
from treating the noises $N_i \ , \ i= 1, 2$ as independent; the
bandwidth, $B$, was taken to be equal to one cycle/$10$ years (i.e.
$3.17 \times 10^{-9}$ Hz). In order to calculate the root-mean-squared
gravitational wave response we have assumed waves to be elliptically polarized and
monochromatic, with their wave functions, ($h_+$, $h_\times$),
written in terms of a nominal wave amplitude, $H$, and the two
Poincar\'e parameters, ($\Phi, \Gamma)$, in the following way \cite{AET1999}
\begin{eqnarray}
h_+ (t) & = & H \ \sin(\Gamma) \ \sin(\omega t + \Phi) \ ,
\label{h2}
\\
h_\times (t) & = & H \ \cos(\Gamma) \ \sin(\omega t) \ .
\label{h3}
\end{eqnarray}
We averaged over source direction and polarization states by assuming
uniform distribution of the sources over the celestial and
polarization spheres respectively. The averaging was done via Monte Carlo
integration with $10000$ source position/polarization state pairs per
Fourier frequency bin and $5 \ \times 10^{5} $ Fourier bins across
the ($10^{-9} - 10^{-6}$) Hz band.
\begin{figure}
\centering
\includegraphics[width=7.0in]{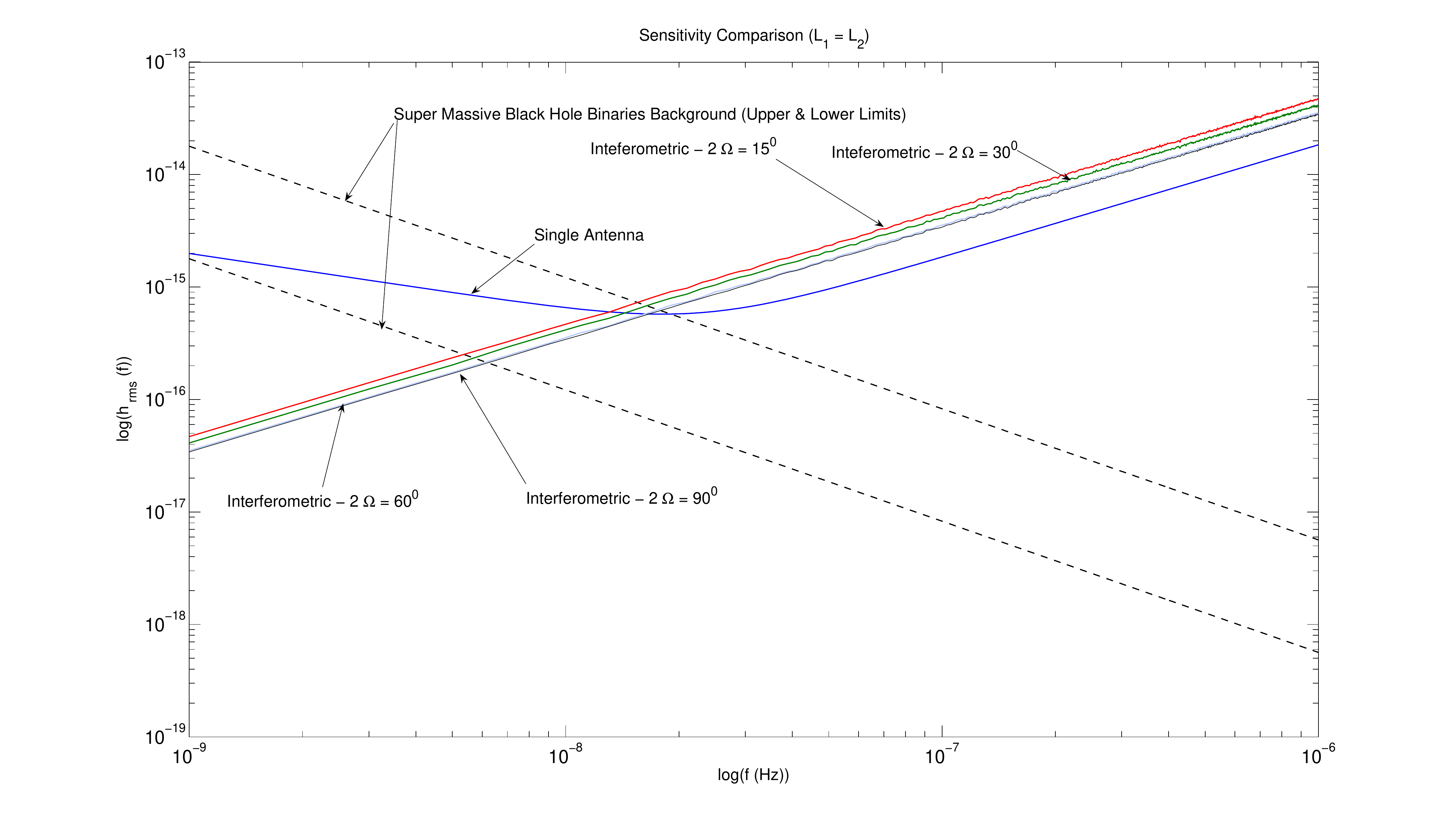}
\caption{Sensitivity of the interferometric
  pulsar timing technique as a function of the Fourier frequency and
  for various values of the pulsars angular separation in the sky, $2
  \Omega$. The pulsars have been assumed to be at equal distance from
  Earth ($1$ kpc). Note the mild dependence of the sensitivity over
  the angle $2 \Omega$ in the interval of values considered. The
  sensitivity of a single-telescope experiment (blue curve), as well
  as the current estimates of the upper and lower bounds on the
  characteristic amplitude radiated by an ensemble of super-massive
  black-hole binaries (two black dashed-lines) are included. See text
  for details.}
\label{fig2}
\end{figure}
In Figure (\ref{fig2}) we plot the sensitivity of the interferometric
pulsar timing technique as a function of the Fourier frequency and for
various values of the angle $2 \Omega$ enclosed between the directions
to the two pulsars. For sake of simplicity we have assumed the two
pulsars to be at an equal distance of $1$ kpc from Earth. We have
independently analyzed configurations with unequal distances and found
no significant differences to the sensitivity curves shown in
Fig.(\ref{fig2}). Another property of the sensitivity curves is their
mild dependence over the angle $2 \Omega$; a configuration with two
millisecond pulsars that are separated by only $15^0$ in the sky shows
a sensitivity that is only $28$ percent worst than that with $2 \Omega
= 90^0$. However, for values of $2 \Omega$ smaller than $15^0$, the
sensitivity rapidly worsen and goes to zero at $2 \Omega = 0$. This is
of course consequence of having assumed, for this calculation, the two
pulsars to be at equal distance from Earth. In the case when $L_1 \neq
L_2$ instead, the sensitivity to gravitational radiation does not go
to zero at $2\Omega = 0$, as it can be easily understood by simple
inspection of Eq. (\ref{Interfero}). Since longer joint tracking times
can be achieved with pairs of millisecond pulsars that have smaller
angular separation, our analysis indicates that these configurations
(with, let's say $2\Omega \simeq 20^0$) might be preferable to those
with larger angle $2\Omega$.

For comparison, in Fig.(\ref{fig2}) we have included the sensitivity
of a single-telescope experiment that relies on a LITS clock
\cite{JAT2011}.  Although at frequencies larger than about $10^{-8}$
Hz this is about a factor of $2$ better than that of the
interferometric technique (in this part of the band the noises
affecting the interferometric combination are twice as many), at
frequencies lower than $10^{-8}$ Hz the advantage of the
interferometric technique becomes evident. In this region of the band
the clock noise dominates all other noise sources in a
single-telescope timing residuals data, and our technique exactly
cancels it.

In order to put in perspective the sensitivity enhancement brought by
the interferometric pulsar timing technique over single-telescope
experiments, we have included the current estimated range of strengths
of an astrophysical gravitational wave background from incoherently
radiating super-massive black hole binaries \cite{Black1,Black2}. At
$3 \times 10^{-9}$ Hz we find that the most conservative prediction of
the strength of such a background would result into an ${\rm SNR}
\simeq 5$, while a more optimistic estimate of its characteristic
strain amplitude would make it detectable with a ${\rm SNR} \simeq
50$.

\section*{Acknowledgments}

I would like to thank Dr. John W. Armstrong for many useful
conversations about pulsar timing, and for his continuous encouragement
while this work was done. This research was performed at the Jet
Propulsion Laboratory, California Institute of Technology, under
contract with the National Aeronautics and Space Administration.(c)
2008 California Institute of Technology.  Government sponsorship
acknowledged.

\end{document}